# On-the-Fly Coding to Enable Full Reliability Without Retransmission

Jérôme Lacan
Université de Toulouse
DMIA-ISAE, LAAS-CNRS
Toulouse, France
jerome.lacan@isae.fr

Emmanuel Lochin
Université de Toulouse
DMIA-ISAE, LAAS-CNRS
Toulouse, France
emmanuel.lochin@isae.fr

***Abstract*—This paper proposes a new reliability algorithm specifically useful when retransmission is either problematic or not possible. In case of multimedia or multicast communications and in the context of the Delay Tolerant Networking (DTN), the classical retransmission schemes can be counterproductive in terms of data transfer performance or not possible when the acknowledgment path is not always available. Indeed, over long delay links, packets retransmission has a meaning of cost and must be minimized. In this paper, we detail a novel reliability mechanism with an implicit acknowledgment strategy that could be used either within these new DTN proposals, for multimedia traffic or in the context of multicast transport protocols. This proposal is based on a new on-the-fly erasure coding concept specifically designed to operate efficient reliable transfer over bi-directional links. This proposal, named Tetrys, allows to unify a full reliability with an error correction scheme. In this paper, we model the performance of this proposal and demonstrate with a prototype, that we can achieve a full reliability without acknowledgment path confirmation. Indeed, the main findings are that Tetrys is not sensitive to the loss of acknowledgments while ensuring a faster data availability to the application compared to other traditional acknowledgment schemes. Finally, we pave the first step of the integration of such algorithm inside a congestion controlled protocol.**

## I. Introduction

In the today Internet, TCP enables by default the Selective ACKnowledgement (SACK) [16] mechanism to perform reliability and the TCP/SNACK variant [7] is preferably used over long-delay links such as satellite links. In the context of DTN, the store and forward nature and the potential disruption events inherent to this architecture might prevent the use of TCP. Indeed, as emphasized in [21], TCP is useful for end-to-end communications across a shared Internet where fairness remains the key parameter and not for store-and-forward communications across a single unshared link where TCP is known to obtain disappointing performances [3]. However, shared satellite links are also used in the context of broadband satellite systems (such as IPSTAR [1] or WildBlue [2]) and reliability is obviously a need in the context of bulk data transfer.

The use of a reliability mechanism based on retransmission can strongly be counterproductive in a DTN context. This is one of the main reason that pushed DTN protocols' designer to see in the Forward Error Coding (FEC) mechanism the best trade-off to implement reliability [21]. However and to the best of our knowledge, we claim that it does not exist a generic reliable mechanism able to efficiently perform over the large range of networks currently enabled. In this study, we rethink the way to enable a generic full-reliability mechanism without introducing overhead that could be counterproductive over various networks characteristics. Following this point, we introduce in this paper a novel erasure coding concept, which realizes on-the-fly coding and which remains resistant to acknowledgement losses, in order to implement a full-reliability algorithm able to perform indifferently over all the networks characteristics. Indeed, the main goal of this erasure code concept is to propose a unified reliability mechanism able to perform over a networks ranging from long-delay networks such as shared satellite, high BER links or standard terrestrial and wireless networks.

The paper is organized as follows: section II presents the context of this study and section III details our proposal. Then, section IV gives a preliminary validation of the proposal while section V provides some analytical tools to benchmark our algorithm. Section VI provides some implementation issues and finally section VII concludes this work.

## II. The big picture

Most of erasure codes (also called FEC.) used over packet erasure channels are block codes [13]. This means at the encoder side, a set of repair packets ($R$) is built from a given set of source data packets (called after source packets and noted $P$) and at the decoder side, these repair packets can only be used to recover source packets from their corresponding set. If too many packets (among the source and repair packets) are lost during the transmission, the recovery of the missing source packets is then not possible. On the opposite, if only few packets are lost, some of the repair packets become useless. A solution to this problem, known as Hybrid FEC-ARQ (or H-ARQ) mechanism [13], is to use receiver's feedback to send additional repair packets or to adjust the redundancy level of the FEC to the observed packet loss rate. However, in a DTN context, large RTT can lead to very long delays to recover efficiently a packet.

The proposal in [15] is to reduce the decoding delay by using non-binary convolutional-based codes. The principle behind is that each repair packets is generated from a sliding window in the set of source packets. However, this mechanism is specifically defined for real-time applications and cannot be directly applied in a DTN context. As an example, it does not provide full reliability service and does not integrate the receivers' feedbacks.

Our proposal can be considered as a mix of these different solutions. The main idea is to build the repair packets from a source packets window (the coding window) which is updated with the receiver's feedbacks. This update is done in a way that any source packets sent is included in the coding window while the sender does not receive any acknowledgement. The method used by the sender to generate a repair packet is simply a linear combination over a binary or non-binary finite field of the data source packets belonging to the coding window while the receiver performs the inverse linear combinations to recover the missing source packets from the received data and repair packets. We allow the user the choice of the coefficients as a trade-off between the best performance (with non-binary coefficients) and the system constraints (the user might prefer the use of binary codes in an energy constrained environment).

One of the main motivation of this study is to propose a reliability algorithm tolerant to acknowledgements loss. We define this property as follows:

*Definition 1:* We define as *tolerant to acknowledgement losses* a mechanism which does not need timely updated information from the receiver to determine which packets must be resent or which coding level must be chosen without impacting on the data availability at the receiver side.

This property has a strong impact in terms of data availability as it avoids the implementation of retransmission timers [17] which are known to be the root cause of spurious retransmission [18]. Indeed, the spurious timeout occurs when a non lost packet is retransmitted due to a sudden RTT increase (handover, high delay variability, re-routing, ...) which implies an expiration of the timer set with a previous and thus outdated RTT value. Several research work have tackled this problem in the context of TCP [10], [14].

In the next section, we detail the algorithm used.

## III. Proposal overview

We denote $P_n$ the $n^{th}$ packet sent (with $P_1$ is corresponding to the first packet) and $R_i^j$ a repair packet for all in sequence packets ranging from $P_i$ to $P_j$. This redundancy ratio is defined as the proportion of repair packets among the total number of packets sent. This ratio can be defined either by the application (following quality of service requirements) or by a cross-layer mechanism (following network characteristics such as BER). Eventually, this number might be dynamically changed during the data transfer.

Let's start with an example: we suppose $4$ packets sent and the fourth one be a repair packet noted: $R_1^3$. This repair packet is computed as follows: $R_i^j = \sum_{k=i}^{j} \alpha_k^{(i,j)}.P_k$ where the $\alpha_k^{(i,j)}$ belong to a finite field fixed. Then, assuming the sender emits packets: $(P_1, P_2, P_3, R_1^3)$; the repair packet $R_1^3$ allows to rebuild a lost packet among three.

Now, let's assume the sender transmits the five packets sequence: $(P_1,\ P_2,\ R_1^2,\ P_3,\ R_1^3)$. If $(P_2{,}P_3)$ are lost, the remaining packets allow to rebuilt the whole sequence. If $(P_1,\ P_2)$ are lost, $P_3$ can be first "subtracted" from the repair packet $R_1^3$ by computing $R'_1 = R_1^3 - \alpha_3^{(1,3)}.P_3$. We then have $(R_1^2, {R'}_1^3)^T = G.(P_1,\ P_2)$ where $G$ is the following $2 \times 2$-matrix:

$$G = \begin{pmatrix} \alpha_1^{(1,2)} & \alpha_2^{(1,2)} \\ \alpha_1^{(1,3)} & \alpha_2^{(1,3)} \end{pmatrix} \tag{1}$$

Clearly, rebuilding $(P_1,\ P_2)$ from $(R_1^2, {R'}_1^3)$ can be done if and only if $G$ is invertible since if $G^{-1}$ exists, we have $(P_1, P_2) = G^{-1}.(R_1^2, {R'}_1^3)$. To improve the probability of having a invertible matrix, the matrix used by the encoder can be extracted from structured matrices like Cauchy [4] or Vandermonde-based [12] matrices. Furthermore, it can be noted that with random matrices, it can be shown that $G$ has an extremely high probability of being invertible if the finite field is chosen sufficiently large [11].

This property leads to efficient transmission and lighten acknowledgement strategies. A broader example is given in Fig. 1. The right side of this figure shows both windows handled by the receiver, one for the repair packets and one for the received packets.

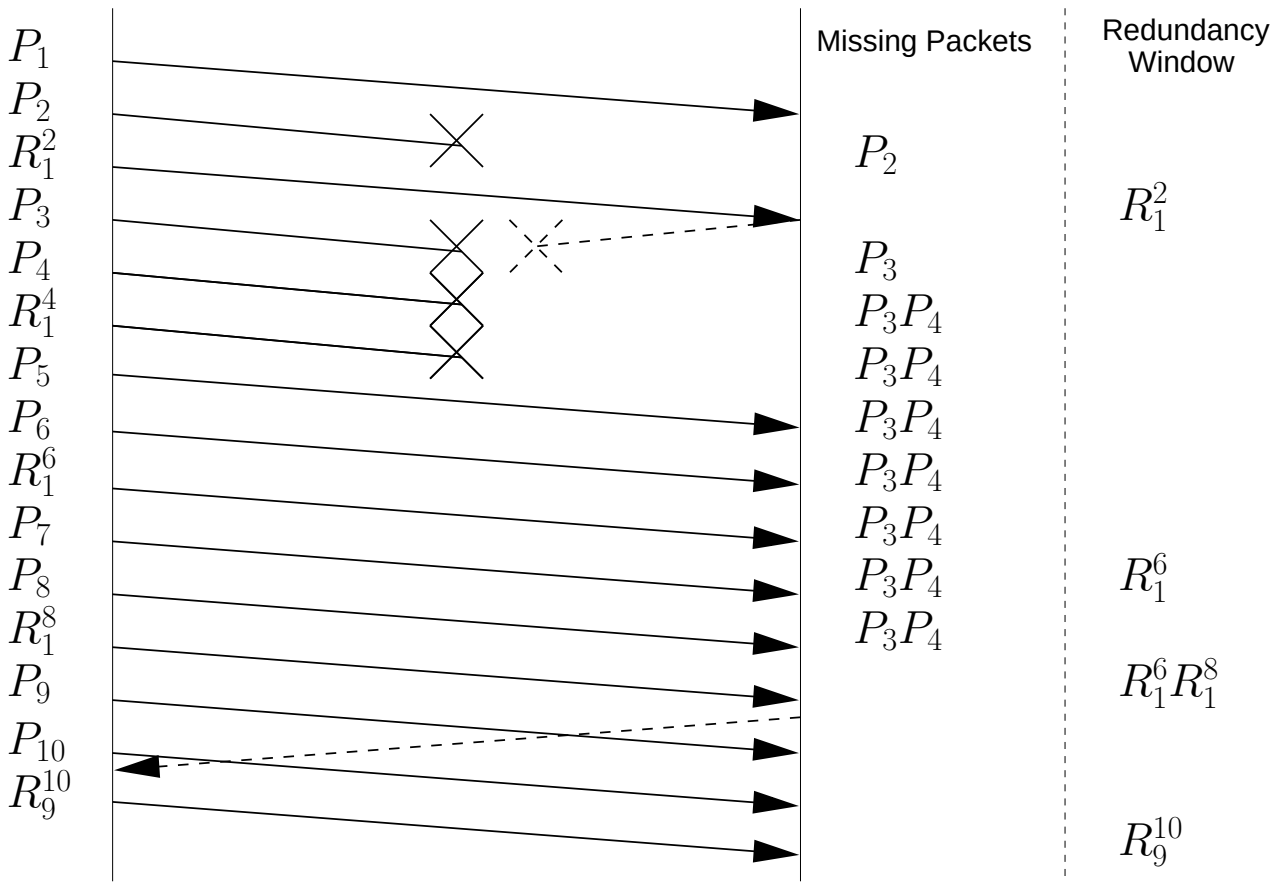

Fig. 1. Reliability mechanism illustration

In this example, during the data exchange, packet $P_2$ is lost. However, the repair packet $R_1^2$ successfully arrived allows to rebuild $P_2$. Then, the receiver sends an acknowledgement packet (dashed line) to inform the receiver that it can compute the next redundancy packets from packet $P_3$. Then, this acknowledgement is lost; however, this lost does not impact on the future of the transmission as the sender still computes a redundancy packet from $P_1$. After this, we can see that $P_3, P_4$ and $R_1^4$ packets are lost. None of these packets need to be retransmitted as they are rebuilt thanks to the packets received from $P_5$ to $R_1^8$. Indeed, as for matrix (1), we can also rebuild $P_3, P_4$ by firstly "subtracting" all the received source packets from the repair packets in order to obtain $(R'^6_1, R'^8_1)$ as follows:

$$(R'^6_1, R'^8_1)^T = \begin{pmatrix} \alpha_3^{(1,6)} & \alpha_4^{(1,6)} \\ \alpha_1^{(1,8)} & \alpha_4^{(1,8)} \end{pmatrix} (P_3, P_4) \qquad (2)$$

Then, the recovery of $P_3, P_4$ can be done by computing the inverse of the $2 \times 2$-square matrix and by multiplying it by $(R'^6_1, R'^8_1)$.

Finally, the number of packets that can handle a repair packet $R_i^j$ depends on the size of the finite field chosen. The larger is this number, the more is the computation complexity. In the case where the use of an acknowledgement path is possible, the dashed acknowledgement line allows to reset this counter as illustrated in Fig. 1. Indeed, this acknowledgement packet function is twofolds: first, it allows to reset the repair counter in order to decrease the computation time (in Fig. 1, after receiving the acknowledgement packet, the repair packet $R_9^{10}$ starts from 9 and not from 1) and second, this confirms the sender that the packets sequence ranging from $P_1$ to $P_8$ are well-received. So, we can note that our mechanism does not depend on the acknowledgements received and is resistant to acknowledgements lost. Indeed, without any feedback received, the sender would send $R_1^{10}$ without any impact on the communication reliability.

## IV. Concept validation

In this section, we validate the concept of this protocol by simulating a bulk data transfer over a lossy link without acknowledgement path. Then, as the packets lost recovery is strongly linked to the success of the matrix inversion process previously explained, we quantify the feasibility of this principle by presenting some matrix inversion statistics obtained by simulation.

We have implemented a prototype of our proposal and used the Linux Network Emulator (Netem) to implement the network losses and delay. We use the same redundancy ratio scheme (equivalent to 33%) as in section III (*i.e.* one redundancy packet sent each two source packets sent). The percentage of losses used during the experiment is ranging from 10% to 33% and the one way delay is set to 50ms (equivalent to an initial RTT of 100ms) and the packets are randomly dropped. We discuss the size of the loss bursts in section VI. All packets have a fixed size of 1040 bytes and each next sent packet is spaced by one millisecond. The total number of packets (*i.e.* data and redundancy packets) sent during each experiment is 3000.

In order to assess the benefit of our proposal, we propose to compare our mechanism with a similar FEC encoding process. For this, we choose a FEC algorithm which create one redundancy packet every two packets sent.

We have to emphasize the difficulty of meaning when comparing our proposal with a FEC or with a retransmission scheme (*e.g.* ARQ) as the goals between all these mechanisms strongly diverge. Indeed, our proposal attempts to implement a reliability mechanism without using packets retransmission through an erasure code which remains non-sensitive to the losses that might occur on the acknowledgement path. On the contrary, the goal of a FEC encoding mechanism is mainly to improve the link transmission quality while a retransmission scheme strictly focus on the success of the data transfer instead of the link occupancy or characteristics. In a DTN context, the DTN protocols recently proposed raise strong hypothesis on the link state in order the transfer succeed as a feedback path is not always available. Then, in this case, our proposal allows these protocols to use a mechanism allowing them to guarantee a kind of reliability of the data transfer. In order to clearly distinguish the goals of each schemes and point out the characteristics of our proposal, we list table I the characteristics of these different mechanisms. The Hybrid-ARQ mechanism (denoted H-ARQ) can be seen as the direct competitor to our proposal. However, H-ARQ is not strictly speaking tolerant to ack losses (following definition 1 in section II). Indeed with H-ARQ mechanism, when a feedback is missing a retransmission timer is triggered. On the contrary with Tetrys philosophy, no such timer must be implemented. This strong choice of design is crucial as most of the retransmission timers used as somehow counterproductive in the context where networks conditions are constantly changing (as previously pointed out in section III).

| | Tetrys | H-ARQ | FEC | ARQ |
|---|---|---|---|---|
| Enable full reliability | YES | YES | NO | YES |
| Improve transmission quality | YES | YES | YES | NO |
| Real time transmission | YES | YES | YES | NO |
| Delay tolerant | YES | YES | YES | NO |
| ACK losses tolerant | YES | NO | N/A | NO |

TABLE I
Comparison between schemes

Table II provides the results obtained by our simulation experiments. This table returns the data packets and redundancy packets lost during the experiments with a lossy channel of 10%, 20%, 30% and 33% packet loss rate. The *total packets lost* line allows us to verify the effective percentage of lost packets. As an example, we can see for the 33% experiment that the real ratio of dropped packets is slightly above this value. Compared to the FEC scheme, we can see that our proposal rebuilt all lost packets when we are strictly below 33% of losses. With the 30% experiment, the 3.15% unbuilt packets occur at the end of the transfer which did not provide enough redundancy packets to rebuild the whole file transferred. In Fig. 2, we represent the number of packets to rebuild with the number of redundancy packets available. When the latter is higher or equal to the former, the algorithm is able to rebuild all missing packets. these three experiments show that our proposal is able to ensure a full data transfer reliability. In Fig. 2(a), we can see that the algorithm can solve a square matrix up to 3x3 elements to rebuild 3 missing packets during the 10% experiment. During the 20% experiment, the maximum matrix size has reached 18x18 elements. Finally, the 3.15% missing packets to rebuild are represented Fig. 2(c) where at the end of transfer, the number of packets to rebuild was higher than the number of redundancy packets currently received (around $t = 1400$). It is important to note that during all the transfer between $t = [0 : 1400]$, all packets have been rebuilt as all matrices have been correctly inverted at the receiver side. Although in the context of a video streaming transfer, these missing packets might be not considered; for a data file transfer, this raises a problem.

Obviously, we cannot predict when the matrices are rebuilt at the receiver side, however, we can estimate the peak matrix size the receiver might have to invert. Thus, the aim of such estimation would allow the sender to optimally send specific redundancy packets in order to ensure high probability to decode the final packets sent. Indeed, keeping in mind that we did not use the acknowledgement path during our experiments, a possible solution to this problem is to transmit only redundancy packets at the end of the file transfer process and to wait for a final closing signal packet from the receiver. Although we want to mitigate the impact of the acknowledgement path, this does not concern the establishment and closing connection process. Except in a DTN or video streaming contexts where we could imagine a sender that do not wait for the closing connection procedure and expect anyway a degree of reliability by ensuring enough redundancy data have been sent to allow the receiver to rebuild the whole sequence, in a general context, a closing procedure has to be implemented. In waiting for the connection FIN packet, the sender would keep sending redundancy packets that could rebuild missing packets and block the closing of the connection. This issue is more a protocol engineering problem rather than an erasure coding problem as it deals with the closing connection process; however an estimation method of the peak matrix size is detailed in the section V.

Finally, we show that our proposal does not outperform the FEC mechanism when the packet loss rate is higher than the redundancy ratio (33% in table II). Fig. 2(d) shows that both data and redundancy packets lines diverge. This means that the number of redundancy packets received does not cover the number of lost packets. In another context, this problem might be lead to an inappropriate choice of the redundancy ratio and as explained at the beginning of section III, a dynamic reconfiguration of this ratio is thus possible through a feedback message from the receiver. This result is expected as we do not propose a mechanism to decrease the losses occurring over a link but the reliability of the data transfer (see table I). As for a FEC code, the main hypothesis of such scheme is to set the redundancy ratio as a function of the Packet Error Rate (PER) which characterizes the network.

In the last Fig. 3, we present an evaluation of the matrix inversion probability. These statistics have been collected with simulations performed in similar conditions than the simulations presented in Fig. 2. Although the use of structured matrices can improve these results, it is well known that random matrices have high probability of being invertible provided the finite field is sufficiently large (see *e.g.* [11]). We thus choose to generate coefficients $\alpha_k^{(i,j)}$ of a repair packet with an uniform pseudo-random generator where, for each packet, the seed is fixed from the timestamp of the packet. With this method, the sender and the receiver can determine all the linear coefficients from the timestamp of the packet.

In the presented simulations, we used the finite field $\mathbb{F}_{256}$, where the elements can be represented with bytes. The results of Fig. 3(b), which presents the respective percentage of successfully inverted matrices, confirm the theoretical results and prove that the matrices are invertible with an extremely large probability (greater than 99% in any case). This means that the recovery is possible as soon as the number of received repair packets reaches the number of lost data packets.

| PLR | 10% | 20% | 30% | 33% |
|---|---|---|---|---|
| PKTS STATS (%) | | | | |
| Total pkts lost | 10.51% | 20.40% | 28.80% | 34.13% |
| Data pkts lost | 12,42% | 26.74% | 37.51% | 50.26% |
| Redundancy pkts lost | 10.41% | 23.45% | 46.72% | 55.03% |
| PKTS NOT REBUILT (%) | | | | |
| Our proposal | 0% | 0% | 3.15% | 99.99% |
| FEC-like | 12.84% | 25.35% | 37.36% | 41.55% |

TABLE II
EXPERIMENTS OVER A LOSSY PATH

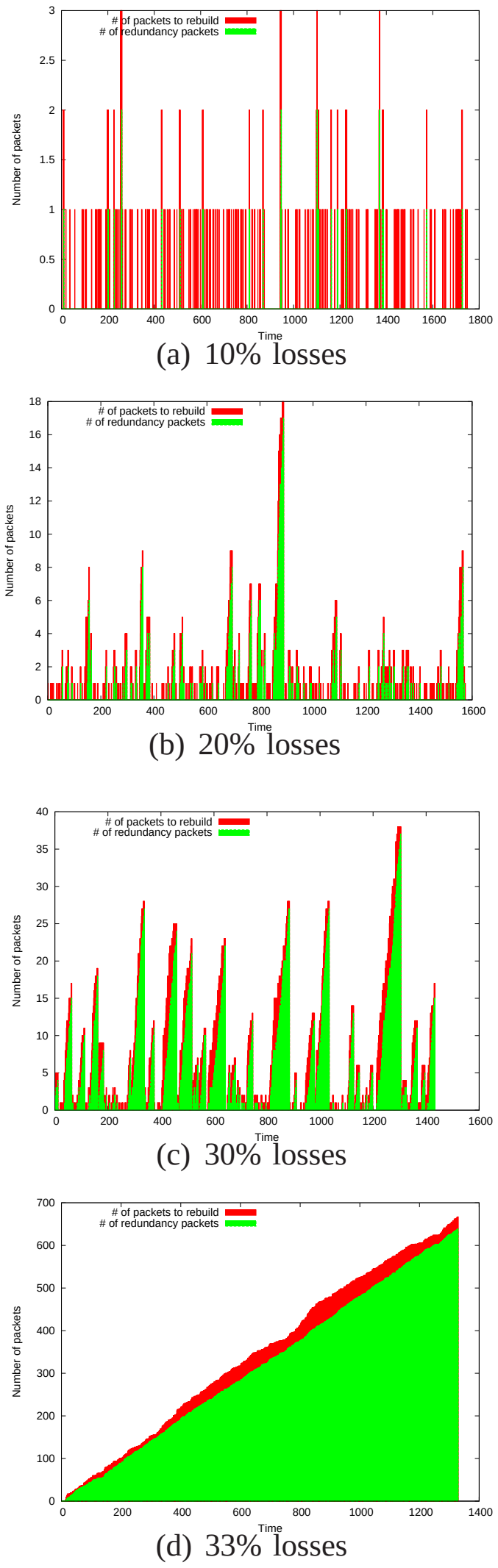

(a) 10% losses
(b) 20% losses
(c) 30% losses
(d) 33% losses

Fig. 2. Number of packets to rebuild and number of redundancy packets received as a function of time during each experiments

Fig. 3(a) gives the number of matrices to invert as a function of the matrices size. This parameter is interesting because the size of the matrix has consequences on:

1) the delay of recovery of the lost data packet and;
2) the amount of computations needed to perform the recovery.

We can observe that the size of the inverted matrix is very small (lower or equal to 2 in most cases) leading to short delay recovery and a low level of computations.

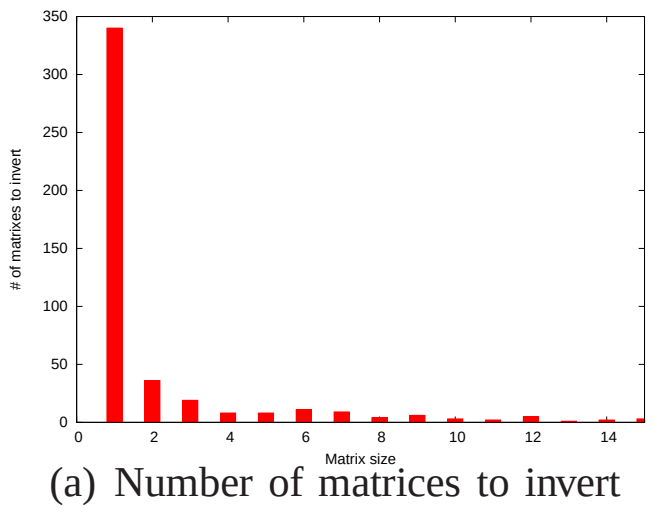

(a) Number of matrices to invert

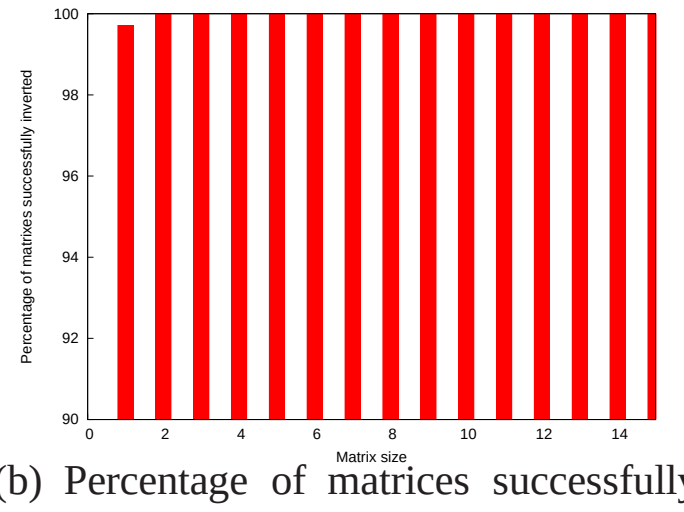

(b) Percentage of matrices successfully inverted

Fig. 3. Statistics on matrices inversion for the 30% experiment

## V. Evaluation of the Mechanism Parameters

This section proposes to evaluate some important parameters in order to assess the performance of our proposal. In particular, we aim at evaluating:

- the time to recover the data. This evaluation allow to compare Tetrys with bi-directional protocols such ARQ or H-ARQ. In this study only Tetrys performances are provided, a comparison study is expected to be driven in a future work with various ARQ and H-ARQ implementations;
- the time between the first loss and the recovery in order to estimate the buffer size the receiver needs;
- the size of the matrices in terms of delay to invert and complexity.

Thus, this section evaluates these different parameters as a function of the parameters characterizing the packet losses of the channel. We consider two kinds of channels. In the first one, we model the packets losses with a Bernoulli model where the losses occur independently following a parameter $p$. Then in the second one, we consider a Gilbert-Elliott model based on a first-order 2-states Markov chain [9] [8]. The "bad" state corresponds to a loss and the "good" state corresponds to a reception of a packet. Despite other more accurate models such as [22], the ones used are general enough to cover a large real part and are commonly used to determine the general behavior of reliability mechanisms in bursty/non bursty channels [5].

The purpose of this evaluation is:

- to compare our proposal with other reliability mechanisms;
- to set the parameters of a mechanism in a given context;
- to define an online strategy to dynamically optimize the parameters in a dynamic environment.

### A. Modeling of the mechanism behavior

*1) Bernoulli channel:* In the first paragraph, we propose a model allowing to estimate the behavior of the mechanism in the case of a Bernoulli model. We consider then that the packets are lost independently with a probability $p$. We denote $r$, the ratio of repair packets among the whole set of sent packets. From the results of Section IV, we assume that a decoding is successful as soon as the number of received repair packets is equal to the number of lost source packets.

We denote $x_i$, the random variable which corresponds to the transmission of the $i^{th}$ packet as follows:

$$X_i = \begin{cases} -1 & \text{if a repair packet is received} \\ 1 & \text{if a source packet is lost} \\ 0 & \text{else} \end{cases}$$

Since the probability of a packet to be a repair packet is independent to its probability to be lost, we have $P(X_i = 1) = r(1-p)$, $P(X_i = -1) = (1-r)p$ and $P(X_i = 0) = (1-r)(1-p) + r.p$.

We then define $Y_i$: the difference between the number of lost source packets and the number of received repair packets:

$$Y_i = \begin{cases} Y_{i-1} + Y_i & \text{if } Y_{i-1} + X_i \geq 0 \\ 0 & \text{else} \end{cases}$$

This system can be seen as an instantiation of the random walk problem extensively studied in the area of stochastic processes. However, most of results on this subject concern the case where $X_i$ has two possible outcomes. In the case of 3 outcomes, the author in [6] studies several cases including those with one reflecting barrier which corresponds to our problem.

Clearly, with this modeling, a decoding is performed at the step $i$ if $Y_{i-1} > 0$ and $Y_i = 0$. Let us introduce the following theorem on the behavior of the mechanism :

*Theorem 1:* If $r > p$, then any lost packet is recovered in finite time. If $r = p$, then any lost packet is recovered, but the mean delay of recovering is not finished.

The first point can be proved by considering that if $r > p$, $P(X_i = 1) < P(X_i = -1)$. The basic results of random walks [6] ensure that the walk returns to $0$ in a finite time. If $p = q$, then $P(X_i = 1) < P(X_i = -1)$ and in this case, it is known that the walk necessarily returns to 0 but the mean decoding time is infinite. Note that if $r < p$, $P(X_i = 1) > P(X_i = -1)$ and the walk ultimately drifts off to $+\infty$, there is no insurance that the packet is recovered.

Now, let us define the random variables corresponding to the parameters we want to evaluate:

- $U$ : the period spent away from the state 0, i.e. the number of consecutive steps where $y_i > 0$. This parameter is called the *recurrence time*. This period corresponds to the time between the first loss after a decoding and its recovery. This time is expressed in time units, where a unit time corresponds to the delay between the transmission of two consecutive packets;
- $T$ : the number of steps where $X_i = 1$ in a period spent away from the state 0. This number corresponds to the number of simultaneously decoded packets. This also corresponds to the size of the matrix inverted by the decoder;
- $D$ : the difference between a step where $X_i = 1$ and the first step where $Y_j > 0$, for $i < j$. This corresponds to the delay between a loss and its recovery. This parameter is called the *decoding delay*.

Following [6], the law of the recurrence time $U$ is characterized by its probability generating function (p.g.f.):

$$\begin{aligned} E(s^U) \;=\; & \Big(1 - s(1-p-r+2.p.r) - \\ & ((1 - s(1-p-r+2.p.r))^2 \\ & -4.p.r.(1-p-r+p.r).s^2)\Big)^{1/2} \\ & \Big/ 2.(1-r).p.s \end{aligned} \quad (3)$$

By evaluating the derivative of $E(s^U)$ for $s = 1$, we obtain the mean $E(U) = \frac{1}{R-p}$. The probability distribution $U$ can be obtained by computing

$$P(U = u) = \frac{1}{u!}\frac{d^u E(s^U)}{ds^u}|_{s=0}$$

Practically, these values can be obtained by expressing $E(s^U)$ as a power series.

From this result, an interesting parameter can be deduced. Indeed, when $Y_i = 0$ (*i.e.* no lost source packet since the last decoding, and thus no repair packet in the buffer), the number of consecutive steps in the state 0 follows a geometric law of parameter $(1 - r)p$. The mean duration of this event is thus equal to $1/(1 - r)p$. From the mean value of $U$, we can deduce the time ratio when the buffer containing repair packets is empty. This ratio is equal to

$$\frac{\frac{1}{(1-r)p}}{\frac{1}{(1-r)p} + \frac{1}{r-p}} = \frac{r-p}{r(1-p)}$$

.

The analytical evaluation of the decoding delay $D$ is not simple. Instead of this exact estimation, we propose a simple approximation of these probabilities. The accuracy of this approximation is illustrated in Figure 4.

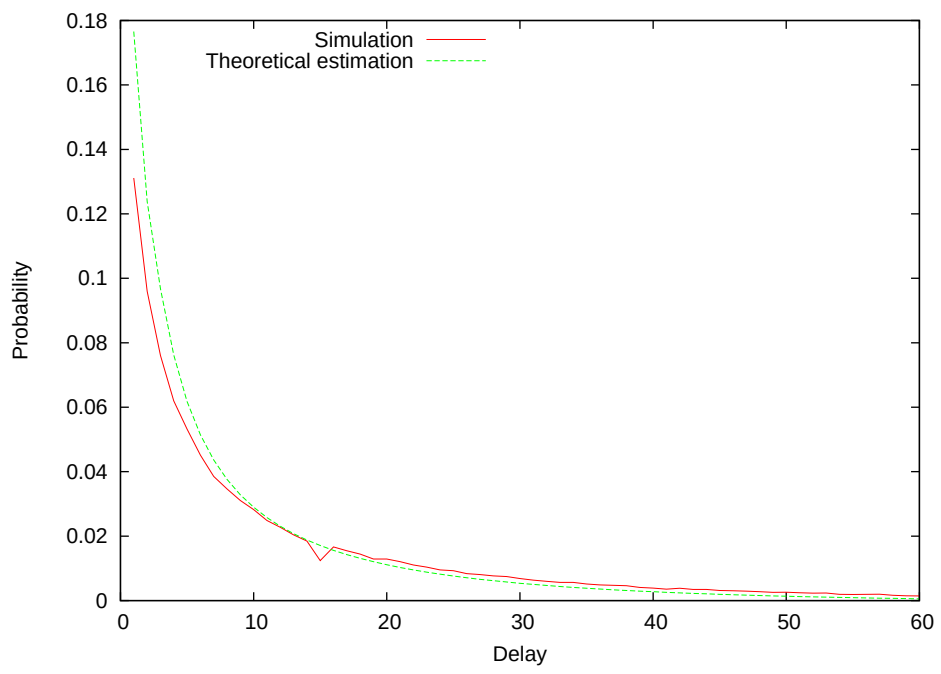

Fig. 4. Comparison between theoretical and simulation results for the decoding delay

In can be noted that the theoretical approximation overestimates the probabilities for small values and underestimates the probabilities for large values. This can be explained by the fact that the approximation assumes that the losses are equally distributed over the time between the first loss after a decoding and its recovery. In reality, this is not the case because it can be observed that the losses have a larger probability to be in the first positions than in the last positions and thus that the real average decoding delay is larger than the approximated one.

To evaluate the sizes of the inverted matrices in the decoding process, *i.e.* the variable $T$, we can consider the joint probability $P(T, U)$ as follows:

$$E(T) \;=\; \sum_{u>t} E(T/U = u)P(U = u) \quad (4)$$

The values of $E(T/U = u)$ can be approximated as follows: first, in the cases where $U = 1$ and $U = 2$, it can be easily verified that $E(T/U = 1) = 1$ and $E(T/U = 2) = 2$. For $u \geq 2$, we can observe that the first step corresponds to a lost packet and the last step corresponds to a repair packet. Between them, there is on average $((1-r)(1-p)+r.p)(u-2)$ packets such that $P(X_i = 0)$. Among the remaining packets, since there are as many received repair packets as lost source packets, we have on average $((1-r)p+r.(1-p))(u-2)/2$ lost source packets. It follows that $E(T/U = u) = 1+((1-r)p+r.(1-p))(u-2)/2$. The mean of $T$ can be then deduced from Eq. 4.

This approximation is defined by introducing the variable $U'$ such that $P(U' = u)$ is the probability that a decoded packet is decoded in the case where $U = u$. Then,

$$P(U' = u_0) = \frac{P(U = u_0).E(T/U = u_0)}{\sum_u P(U = u).E(T/U = u)}$$

We then have :

$$P(D = d) \;=\; \sum_{u>d} P(D = d/U' = u)P(U' = u) \quad (5)$$

Actually, $P(D = d/U' = u)$ is the probability that a decoding delay of $d$ is observed knowing that this packet was decoded in the case where $U = u$. Concerning the approximation of this probability, we propose to consider that there are $E(T/U=u)$ losses in the case where $U=u$. The first position is a lost packet and thus, its decoding delay is $u$. Thus, $P(D=u/U'=u)=\frac{1}{E(T/U=u)}$. We consider that all the other positions have the same probability to be a lost packet. Their approximated probability is thus equal to $(1-\frac{1}{E(T/U=u)})/(u-1)$.

The values of $P(D)$ can be then obtained from Eq. 5.

*2) Markov channel:* This modelling was proposed for the case where the packet losses follows of a Bernoulli law. However, it is well known that in most of the channels, this model is not sufficient. A class of more powerful models is based on Markov chains such as the Gilbert-Elliott model [8], [9].

To evaluate the proposed mechanism over such channels, we consider a simple first-order 2-states Markov model with a bad state corresponding to a packet loss and a good state corresponding to a packet reception. The values $P_1$ and $P_2$ represent respectively the packet loss rate in the good and bad states (see Figure 5). It can be easily shown that the mean packet loss rate of this model is equal to $P_1/(1+P_1-P_2)$ and that the mean burst length is equal to $1/(1-P_2)$.

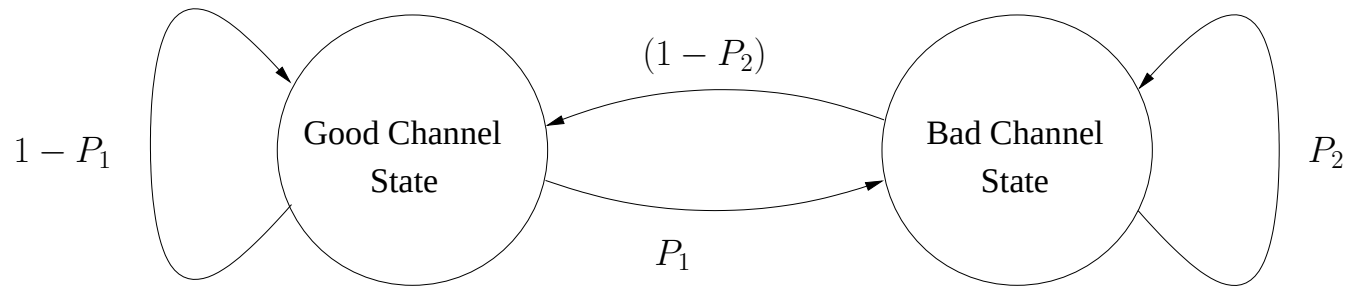

Fig. 5. Good and bad states

Since the modeling in the Bernoulli model was based on a random walk which is also a Markov chain, it is possible to combine these two chains to define a global Markov chain which models the behavior of the mechanism in this type of channel. This chain, represented on Figure 6, is composed of states characterized by the state of the channel (bad/good reception) and by the difference between the number of received repair packets and the number of lost source packets. The transition probabilities are then deduced from the type of packet (source/repair) and by the reception (lost/received).

From this Markov chain, the parameter studied for the Bernoulli channel can also be evaluated using Markov chain theory. This study is planed for future work.

## B. Simulation results and analysis

Several simulations were done in order to evaluate the behavior of the mechanism as a function of the different parameters in the two channel models. For these evaluations, these 2 models were explicitly added to our mechanism implementation in order to generate loss traces following accurately the probabilities laws. For the decoding, a Gauss-Jordan matrix inversion was developed from the Reed-Solomon codec of L. Rizzo [19]. This algorithm was modified in order to determine, in the case of a non-singular matrix, the repair packet which is a linear combination of the other received packets. This packet, which is useless, is then discarded and the decoder waits for additional repair packets. In these simulations, the coefficients for the linear combination are randomly chosen on the finite field $\mathbb{F}_{256}$, except in Section V-B3 where other finite fields are used.

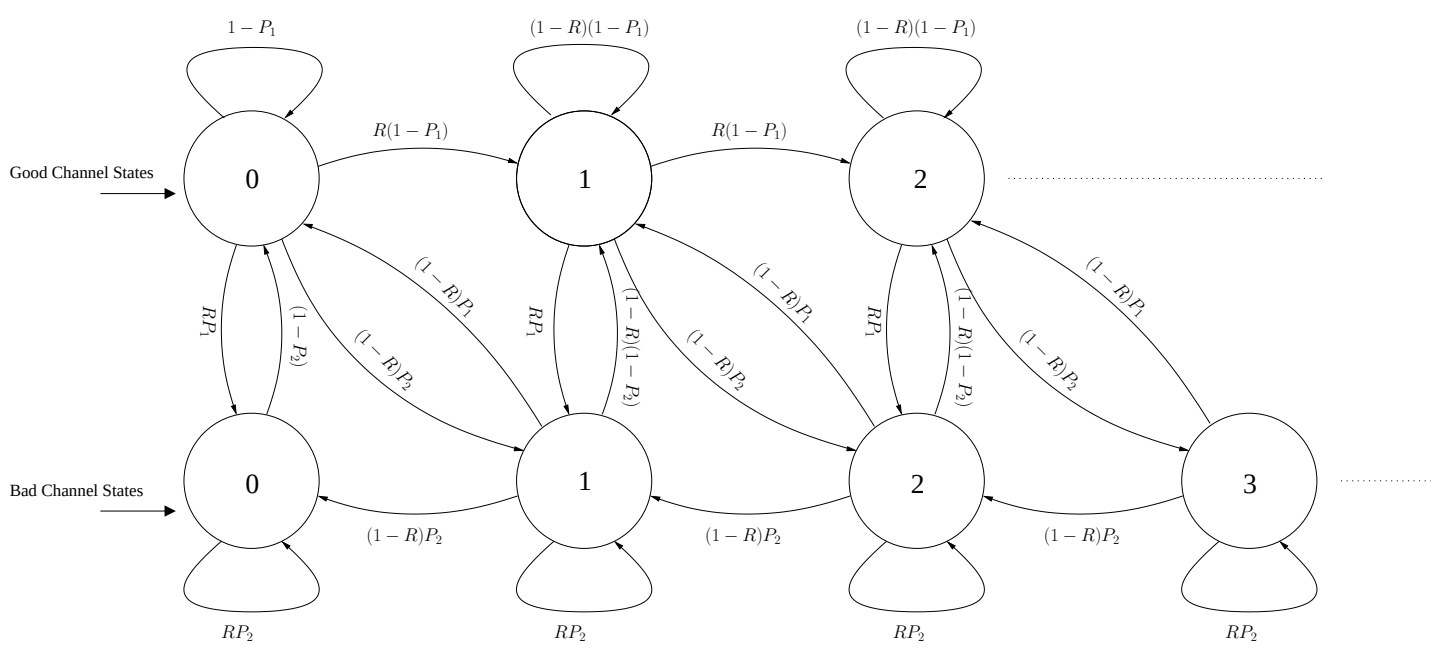

Fig. 6. Markov model

*1) Variable input probabilities:* Figure 7 illustrates the variations of the mechanism performances in terms of mean decoding delay, mean matrix size and mean recurrence time on a Bernoulli channel in the case where the repair ratio is fixed to 0.25 and the packet loss rate $p$.

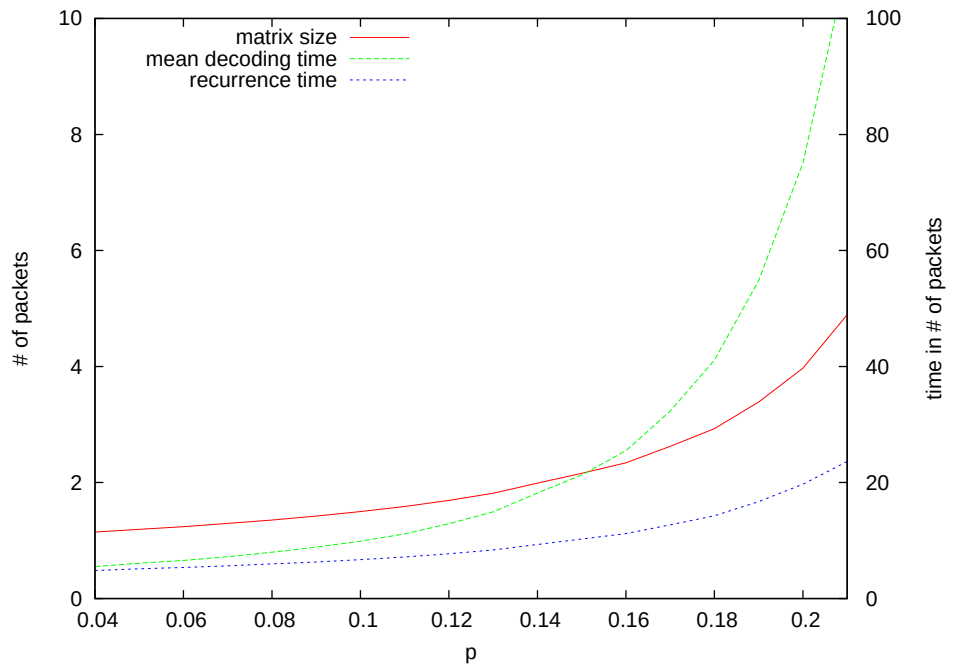

Fig. 7. matrix size, mean decoding time and recurrence time as a function of $p$

The error probability is represented in the x-axis and the matrix size, mean decoding time and recurrence time are represented in the y-axis. We used two scales in the y-axis. The first one (on the left side) is expressed in number of packet and is used for the mean matrix size. The second scale (on the right side) is expressed in time units. Recall that a unit time corresponds to the delay between the transmission of two consecutive packets. This time scale is used for mean decoding time and the recurrence time.

The first observation is that the three curves increase with the packets lost rate. This can be explained by the fact that when the error probability is very small compared to the repair ratio, then the decoding are done very soon and thus, the recurrence time, the decoding delay and the size of the inverted matrices are small. When the packets lost rate grows towards the repair ratio, the three curves increase. It can be recalled from the previous Section that the average recurrence time is equal to $1/(R-p)$ and thus, that is infinite when $R=p$.

We can also observe that the "decoding delay" curve becomes larger than the "recurrence time" curve. This can be explained by the fact that the decoding delay is related to packet, instead of the recurrence time is related to decoding. In the case of a large "recurrence walk", a large number of packets have a large decoding delay, and thus this walk has a larger influence on the average decoding time than on the average recurrence time.

*2) Variable burst size:* Figure 8 shows the influence of the burst of losses. We consider the first-order 2-states channel model of Figure 5. The input loss probabilities $P_1$ and $P_2$ vary in such a way that the mean packet loss rate is kept constant (equal to 0.2). The repair ratio is fixed to 0.25.

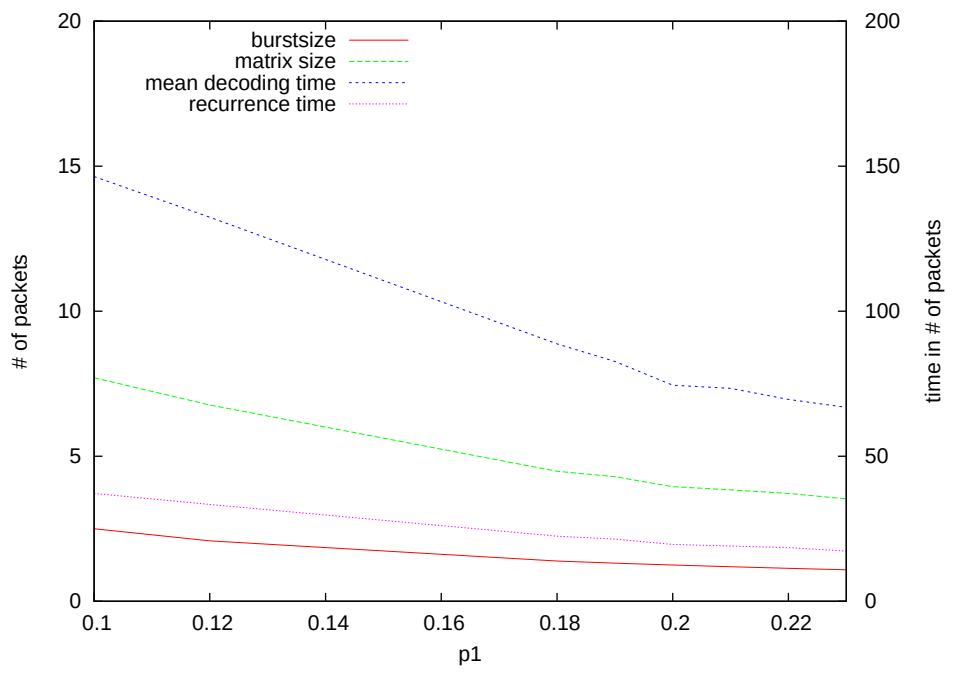

Fig. 8. Impact of the burst size on the matrix size, mean decoding time and recurrence time

The parameter $P_1$ is represented on the x-axis. The value of $P_2$ can be deduced from the packet lost rate $p$ and $P_1$. Indeed, recall from Section V-A2 the $p = P_1/(1 + P_1 - P_2)$. Thus, $P_2 = 1 + P_1 - P_1/p$.

Compared to the previous Figure, the curve representing the mean burst length (equal to $1/(1 - P_2)$) is added. We can observed that a small value of $P_1$ implies a large value of $P_2$ and thus a large mean burst size. On the opposite, when $P_1 = P_2$, the Markov channel becomes a Bernoulli channel of parameter $P_1$ and thus, the mean burst size reaches its minimum.

The main information of the Figure 8 is that the burst losses have a negative impact of the matrix size, mean decoding time and recurrence time. We can observe that when $P_1$ varies from $0.1$ to $0.2$, the burst size varies from $2.5$ to $1.25$. In this range, the matrix size, mean decoding time and recurrence time are also divided by 2.

Even this rate of 2 is very specific to this simple example, more generally, we can observe that the only consequence of bursts is the increase of the decoding delay, recurrence time and of the matrix size at the decoder side. This can be compared, for example, to classical FEC which have to implement inter-leavers or very long codes to cope efficiently with bursts.

Note that in the case of channels with variable parameters (with a fixed packet loss rate), our mechanism adapts naturally to the variable channel without extern intervention.

*3) Variable finite field:* Figure 9 shows the influence of the finite field size on the outputs. Indeed, it was shown in Section III that the decoding is not necessarily possible as soon as the number of received repair packets is equal to the number of lost source packet. This can be explained by the fact that the corresponding matrix is not singular. In this case, the receiver must wait additional repair packets and then the delay and the matrix size are increased. In this simulation, the finite field size (on the x-axis) varies from $2^1$ to $2^8$. Recall that the coefficients used to build the repair packets are randomly chosen.

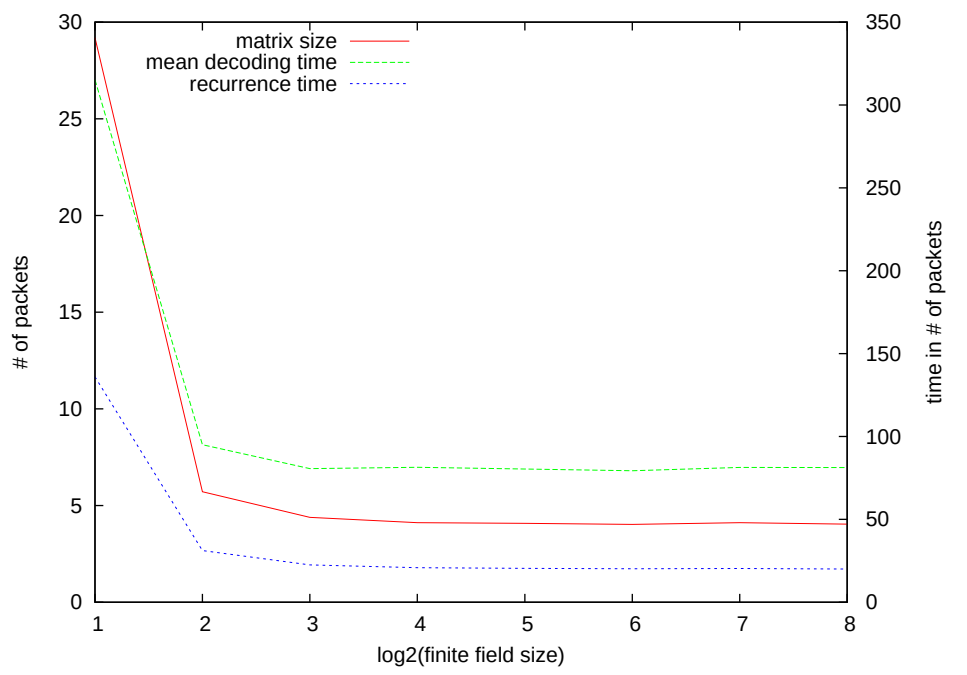

Fig. 9. Impact of the finite field size on the matrix size, mean decoding time and recurrence time

The main result of this figure is that the two smallest finite fields ($\mathbb{F}_2$ and $\mathbb{F}_4$) obtain performance significantly worst than the others finite fields. Even if the binary field is attractive because all its operations can be implemented with extremely fast XORs operations, this field should be avoided in our mechanism. The best compromise seems to be the field $\mathbb{F}_8$ which obtain excellent decoding performance while supporting very fast operations.

Note that the use of non-random binary matrices from efficient binary erasure codes like *e. g.* Raptor codes [20], should reduce the gap between binary and non-binary field. This work is planned for future work.

## VI. Implementation Issues

In the previous sections, we have illustrated through real measurements and simulations the feasibility of the concept proposed. From these results, several points related to the implementation of the mechanism must be discussed.

The first one is the the size of the finite field. This point was presented in the previous Section where it has been shown that the size of a finite field must be at least of $8$.

The second point deals with the loss patterns and the packet loss rate in the channel. It has been shown that when the repair ratio is larger than the packet loss rate, the mechanism correctly performs. However, the decoding delay can be large. In the case of real-time applications, the repair ratio can be tuned in order to obtain decoding delays smaller than the maximum delay acceptable by the application. Note that the packet loss rate and the model of the 2-states Markov channel can be updated during the transmission [22].

Another important point concerns the patterns of repair packets at the sender side. Indeed, in the modelling and the simulations, we have just fixed the repair ratio. Actually, it can be observed that the performance of the mechanism can be significantly improved if the repair packets are sent in a regular interval. For example, for a packet loss rate of $0.1$ over a Bernoulli channel, with repair packets randomly sent with a rate $0.1667$, the mean decoding delays is equal to $33.30$. On the same channel, if the sender repeats the following pattern: one repair packet sent every 5 source packets, the repair ratio is still $0.1667$, but the observed mean decoding delays is equal to $15.88$.

The complexity is also a point that must be carefully evaluated. Two kinds of operations must be performed by the decoder. First, it has to subtract the received source packets to the received repair packets. This operation can be performed on-the-fly as soon as a repair packet is received. Note that the average number of packets that must be subtracted depends on the recurrence time. The second operation is the matrix inversion and the multiplication of the matrix by the repair packets to recover the source packets. For a $m \times m$-matrix, the inversion is done in $O(m^3)$ operations and the multiplication of the matrix by the packets needs $O(sz.m^2)$ operations, where $sz$ is the number of symbols of the finite field in each packet. Clearly, the amount of computations depends on the recurrence time and matrix size. Since these parameters depends on the difference between the packet loss rate and the repair ratio, we can establish a direct relation between the amount of computations and the repair ratio. Consequently, a device constrained in energy can enlarge the repair ratio in order to reduce the amount of computations.

## VII. Conclusion

In this paper, we propose a novel reliability mechanism, based on a new on-the-fly erasure coding, enabling an implicit acknowledgment strategy. To the best of our knowledge, this algorithm is the first one that allows to unify a full reliability service with an error correction scheme. In this paper, we detail the algorithm and provide real measurements to illustrate the behavior of Tetrys. We model the performance of this proposal and demonstrate with a prototype, that we can achieve a full reliability service without acknowledgment path confirmation due to the non sensitivity of Tetrys to the loss of acknowledgments while ensuring a faster data availability to the application. Furthermore, this proposal allows a fast recovery compared to block codes and avoids non-useful retransmitted packets. In a future work, we expect to drive a performances comparison study with a large range of measurements and various network conditions to assess the benefit of our proposal compared to others existing schemes such as H-ARQ.